\documentclass[aps,pra,twocolumn,amsmath,amssymb,showpacs,superscriptaddress]{revtex4-2}

\usepackage[version=3]{mhchem} % Formula subscripts using \ce{}

\usepackage{graphicx} % Required for inserting images
\usepackage{dcolumn}
\usepackage{natbib}
\usepackage{hyperref}
\usepackage{amsmath}
\usepackage{amssymb}
\usepackage{ulem}
\usepackage{xcolor}

\begin{document}

%\title{Printed diffractive optics for orbital angular momentum transfer in optical tweezers}
\title{Robust Orbital Angular Momentum Transfer Using Low-Cost Diffractive Optics}
\author{Beatriz Morales-Cruzado}
\affiliation{SECIHTI - Centro de Investigaciones en Óptica, A.C., Loma del Bosque 115, Colonia Lomas del Campestre, León, Guanajuato C.P. 37150, México}
\author{Benjamin Perez-Garcia}
\affiliation{Photonics and Mathematical Optics Group, Tecnologico de Monterrey, Monterrey 64849, México}
\author{Francisco G. Pérez Gutiérrez}
\affiliation{Facultad de Ingeniería, Universidad Autónoma de San Luis Potosí, Av. Manuel Nava No. 8, C.P. 78290, San Luis Potosí, SLP, México}
\author{Carmelo Rosales-Guzmán}
\affiliation{Centro de Investigaciones en Óptica, A.C., Loma del Bosque 115, Colonia Lomas del Campestre, León, Guanajuato C.P. 37150, México}
\email{carmelorosalesg@cio.mx}

\date{\today}

\begin{abstract}
Reliable transfer of orbital angular momentum (OAM) to microscopic objects typically relies on high-fidelity vortex beams generated by programmable spatial light modulators or precision-fabricated phase optics. Here, we demonstrate that robust OAM transfer in optical tweezers can be achieved using static binary holograms printed on acetate substrates. The printed diffractive optics generate Laguerre–Gaussian vortex beams with sufficient spatial fidelity to induce controlled optical torque and stable rotational manipulation of polystyrene microspheres in a high-numerical-aperture optical tweezers system. Despite a diffraction efficiency of only approximately 2\%, the generated beams enable reproducible particle rotation using less than 1 mW of optical power in the first diffraction order. The rotational dynamics were systematically characterized as a function of incident optical power and topological charge, revealing the expected increase in angular velocity with both parameters, consistent with OAM-driven torque in the overdamped regime. These results demonstrate that efficient optical angular momentum transfer is remarkably tolerant to the reduced efficiency of passive printed diffractive optics, establishing a robust, scalable, and high-damage-threshold platform for structured-light optical manipulation with applications in microfluidics, biophysics, optomechanics, and optical trapping.
\end{abstract}

\maketitle

%\section{Introduction}
Optical tweezers have become a powerful tool for trapping and manipulating microscopic and nanoscopic objects using highly focused laser light \cite{marago2013optical,daly2015optical}. Notably, they have been extended into the plasmonic regime, where surface plasmons are exploited to enhance light--matter interactions and enable trapping functionalities beyond those achievable with conventional optical tweezers \cite{zhang2021plasmonic,min2013plasmonic}. In both cases, trapping arises from optical forces generated by spatial intensity gradients, which draw particles toward regions of maximum field confinement and enable precise, contactless control \cite{gieseler2021optical}. As a result, optical tweezers have become a central tool in fields such as biophysics, physics, and engineering, enabling the manipulation and quantitative analysis of forces at cellular and molecular scales \cite{favre2018optical,zhang2017ultrasensitive,dholakia2006optical}.

A major advance in optical trapping has been driven by the development of structured light, where optical beams are tailored in their various degrees of freedom, such as, phase, amplitude or polarization \cite{woerdemann2013advanced,2021_Yang}. Structured light fields impart additional functionalities to conventional optical tweezers, enabling enhanced control over particle dynamics. Early work in this direction focused on light beams carrying angular momentum, which enable not only stable trapping but also controlled rotational motion of trapped objects. Despite the remarkable capabilities enabled by structured light, their widespread adoption in optical tweezers remains limited by practical considerations. In particular, the use of computer-controlled devices such as spatial light modulators (SLMs), which provide dynamic control over the intensity, phase, and polarization of optical fields, substantially increases system complexity and cost. While SLM-based platforms enable advanced functionalities (including three-dimensional particle manipulation, dynamic beam shaping, and parallel trapping) they often require specialized hardware, high-performance electronics, and nontrivial calibration procedures \cite{SPIEbook}. These requirements hinder the dissemination of structured-light optical tweezers beyond well-equipped laboratories and limit their accessibility in resource-constrained environments. Consequently, there is a growing need for alternative approaches that retain the essential functionalities of structured light while significantly reducing system cost and experimental overhead. Developing low-cost, robust, and easily implementable beam-shaping strategies is critical for broadening the impact of optical tweezers across disciplines such as biophysics, soft matter, and microengineering. In this context, static or quasi-static optical elements capable of generating structured beams offer an attractive pathway toward accessible optical trapping platforms.

In this manuscript, we demonstrate that low-cost binary printed holograms constitute a practical and scalable alternative for generating structured light in optical tweezers applications. By generating orbital-angular-momentum (OAM) beams with printed holographic elements, we achieve controlled transfer of angular momentum and reliable particle rotation without the need for expensive spatial light modulators (SLMs). Importantly, although the transfer of OAM from light to microparticles has been widely demonstrated, the quality of the generated OAM beams remains critical for inducing rotational motion. Inhomogeneous intensity distributions can suppress rotation, as efficient OAM transfer requires a uniform annular intensity profile. Otherwise, localised high-intensity regions act as optical potential wells that trap particles rather than drive their rotation. Moreover, a ring-shaped intensity distribution alone does not guarantee the presence of OAM. For example, a zero-order Bessel beam exhibits an annular intensity distribution, yet lacks the azimuthal phase structure required to carry OAM. A further challenge arises from the optical power required to induce mechanical rotation, which directly depends on the conversion efficiency of the generation method. For low-cost printed holograms, this efficiency is typically limited, which reduces the amount of optical power available for OAM transfer. Together, these factors highlight the difficulty of consistently observing OAM-induced rotation using inexpensive optical elements. Nevertheless, as it will  be demonstrated, the LG beams generated with printed holograms can reliably transfer OAM from light to matter. These results establish printed holograms as an accessible platform for structured-light optical manipulation and open new opportunities for the development of affordable, reproducible, and widely deployable optical trapping systems. 

\begin{figure}
    \centering
    \includegraphics[width=0.9\linewidth]{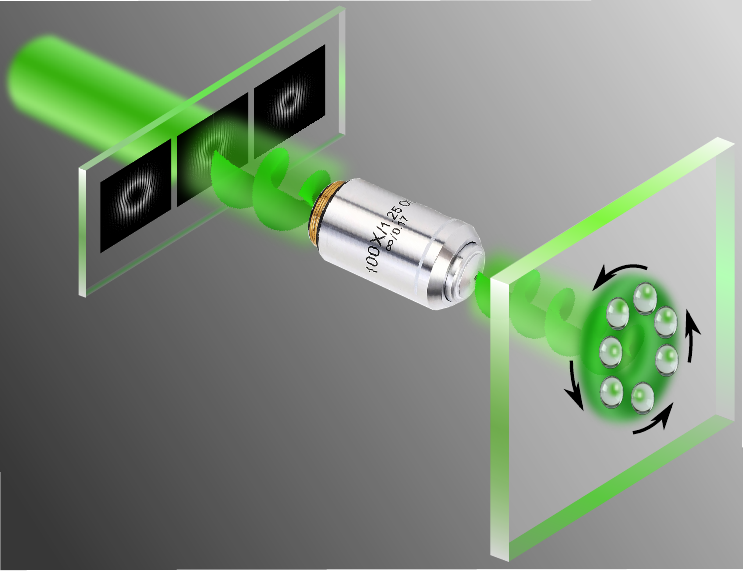}
    \caption{Schematic of the holographic vortex optical tweezers system. Binary holograms printed on acetate generate optical vortex beams carrying OAM, which is transferred to trapped microparticles to induce controlled rotational motion.}
    \label{fig:concept}
\end{figure}

Figure \ref{fig:concept} schematically illustrates the proposed approach, in which a single laser beam illuminates a set of printed fork holograms with different topological charges. Each hologram generates a light beam carrying OAM associated with its topological charge. The holograms are arranged horizontally and mounted on a simple translation stage, enabling rapid and intuitive switching between different OAM beams without the need for active beam-shaping hardware.

%\section{Generation of LG beams with printed binary holograms}

\begin{figure}
    \centering
    \includegraphics[width=0.9\linewidth]{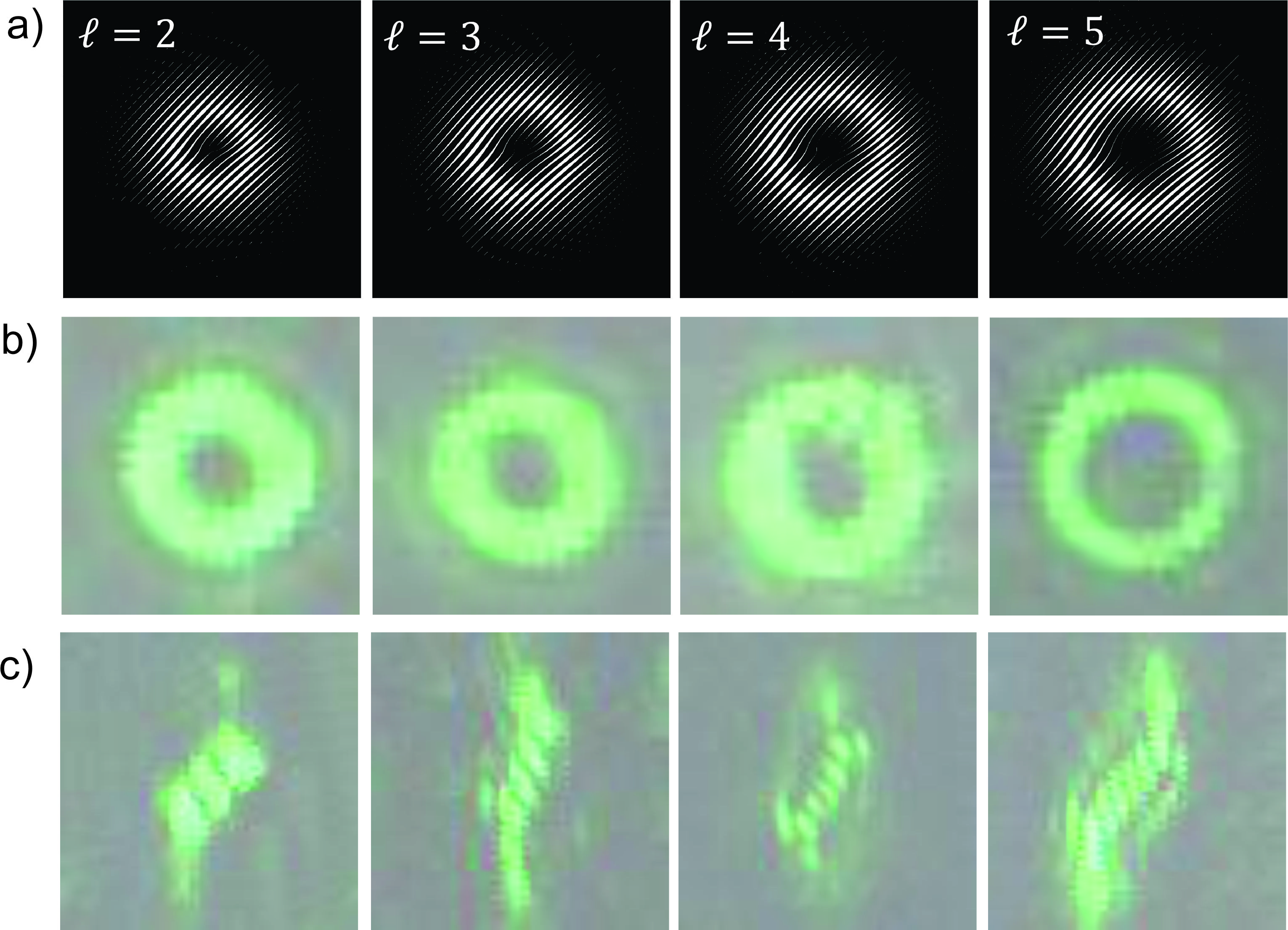}
	\caption{a) Binary holograms, b) the corresponding intensity distributions in the $+1$ diffraction order at the trapping plane of the optical tweezers system, for different values of the topological charge $\ell$, and c) intensity patterns obtained after an astigmatic transformation using a tilted spherical lens.}
    \label{fig:hologramas}
\end{figure}

LG beams carry an OAM of $\ell \hbar$ per photon, determined by the topological charge $\ell$. When such beams interact with microscopic particles, they transfer optical torque that induces rotational motion. A formal explanation can be found in \textbf{Section I of the Supplementary Material}.
In this work, LG beams are generated using printed binary holograms that encode both amplitude and phase information through spatial modulation. The holograms include a carrier grating that separates diffraction orders, enabling the desired vortex beam to be obtained in the first diffraction order with high fidelity. Details of the generation of binary holograms can be found in \textbf{Section II of the Supplementary Material}. The holograms selected correspond to LG beams with $p=0$ and a topological charge $\ell$ ranging from $2$ to $5$. In Fig.\ \ref{fig:hologramas}a) and \ref{fig:hologramas}b), the printed holograms and their intensity distribution in the $+1$ diffraction order are shown. It is observed that the size of the doughnut-shaped light distribution increases as the absolute value of the topological charge increases, resulting in a doughnut of larger radius. The intensity distributions of the LG modes were projected onto the trapping plane of an optical tweezers system. For $\ell =2$, the internal diameter measures approximately $0.44\ \mu\mathrm{m}$ and the external diameter approximately $2.5\ \mu\mathrm{m}$, while for $\ell=5$ these diameters become $1.03\ \mu\mathrm{m}$ and $3.08\ \mu\mathrm{m}$, respectively. The OAM content of the generated beams was verified using the astigmatic transformation method introduced by Vaity and Rusch \cite{vaity2013measuring}. This technique employs a tilted spherical lens to transform the helical phase structure of the LG beam into a Hermite-Gaussian-like intensity distribution, as shown in Fig. \ref{fig:hologramas}c). The topological charge is then determined from the transformed pattern, where the absolute value of the charge corresponds to the number of intensity lobes minus one. Furthermore, the orientation of the lobes reveals the sign of the topological charge, a positive inclination corresponds to a positive topological charge, whereas a negative inclination corresponds to a negative topological charge. 

%\section{Experimental methodology}

\begin{figure}
    \centering
    \includegraphics[width=0.9\linewidth]{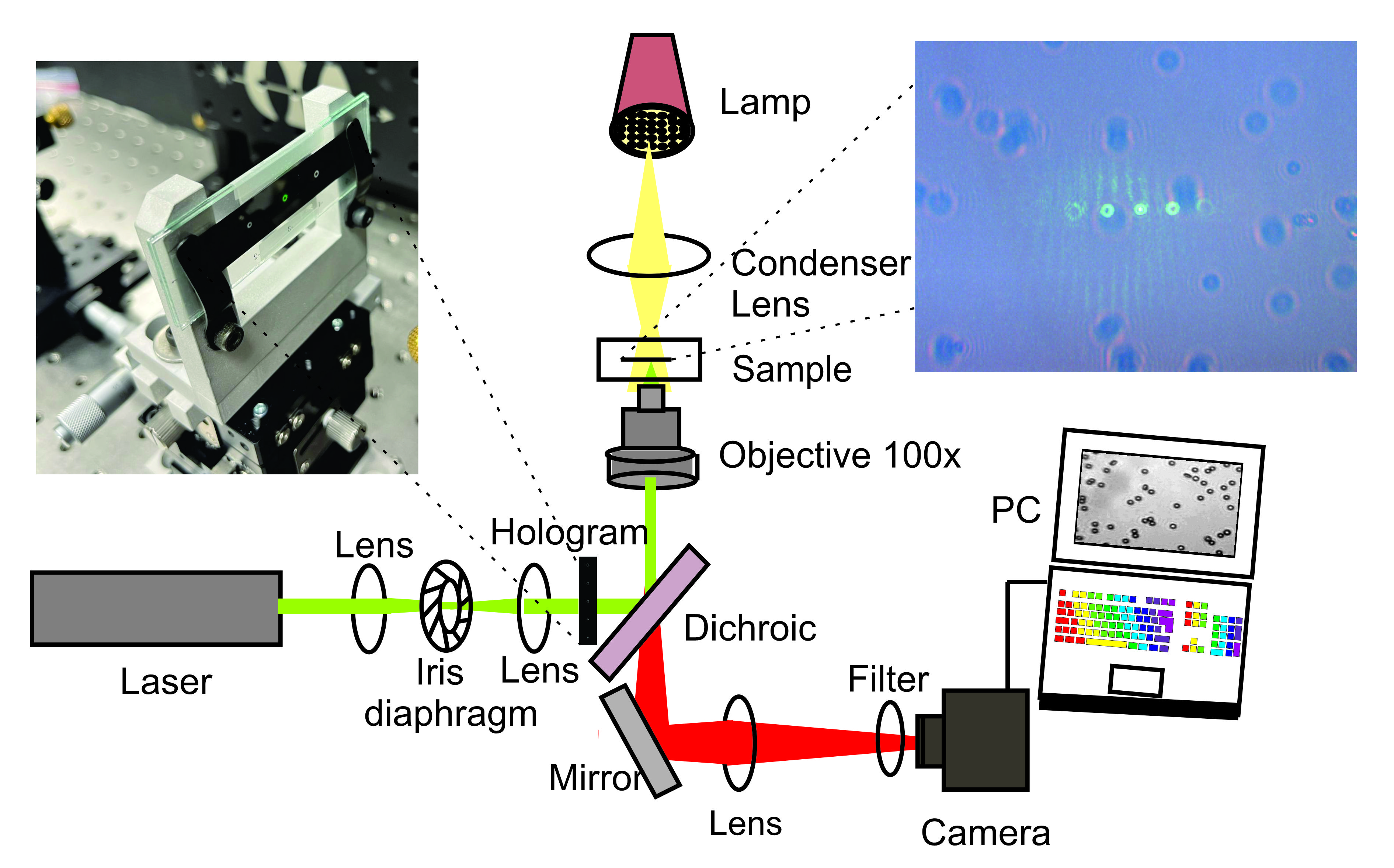}
    \caption{Experimental setup of the vortex holographic optical tweezers system.  The inset on the left shows a photograph of the acetate-printed hologram (shown in better detail in figure \ref{fig:hologramas}), and the photo on the right shows an image at the trapping plane where the distribution of light in the different diffraction orders can be observed.}
    \label{fig:experiemntalsetup}
\end{figure}

To investigate the transfer of OAM, an optical tweezers system based on printed binary holograms was implemented. A continuous-wave laser at 532 nm is expanded and directed into an optical tweezers system with a high-NA objective (NA = 1.25). Printed holograms are inserted before the objective to generate LG beams at the trapping plane as shown in the figure \ref{fig:experiemntalsetup}. Details of the complete experimental setup can be found in \textbf{Section III of the Supplementary Material}.

A series of acetate-printed holograms were inserted after the telescope, at a distance of $8.0$ cm from the back of the microscope objective to generate LG beams with radial index $p=0$ and topological charge $\ell$ ranging from $2$ to $5$. The holograms were mounted on a translation stage that facilitates the transition between the different topological charges. The top-left inset of Fig.\ \ref{fig:experiemntalsetup} shows a photograph of the binary hologram, where the binary structures formed by transparent and dark regions can be observed. As explained before, these acetate-printed holograms modulate the incident beam in amplitude and phase to generate the desired LG beam. Furthermore, since they act as a diffractive element, different diffraction orders emerged, each containing repetitions of the structured beam. In particular, the $+1$ and $-1$ diffraction orders contained OAM beams with equal topological charge magnitude but opposite signs. The generated LG beams, featuring a doughnut shape, were then projected onto the trapping plane, as illustrated in the top-right inset of Fig.\ \ref{fig:experiemntalsetup}, where the transverse intensity distribution of various diffraction orders is observed. 

Polystyrene microparticles ($1\ \mu\mathrm{m}$ and $2\ \mu\mathrm{m}$  diameter) suspended in water are trapped within the annular intensity profile of the vortex beams. The dependence of the particle's angular rotation on the input laser power and topological charge was systematically investigated. Specifically, the optical power incident on the printed holograms was varied from $0$ to $400\ \mathrm{mW}$, while the topological charge $\ell$ was tuned between $2$ and $5$. For each combination of particle size, laser power, and topological charge, a video sequence was recorded. These videos were subsequently processed to extract the corresponding angular velocity of the rotating particles, as detailed in \textbf{Section IV of the Supplementary Material}. 

As discussed above, the generated LG beams carry an OAM of $\ell\hbar$ per photon, which can be transferred to trapped microparticles, resulting in their rotational motion. However, the mere presence of OAM does not guarantee observable rotation, since small deviations from the ideal LG intensity or phase profile can severely limit the efficiency of OAM transfer. Here, it is shown that despite being produced with low-cost optical elements, the generated beams not only robustly impart rotational motion to microparticles but also yield angular velocities that scale linearly with the topological charge. 

To begin with, it is important to note that the conversion efficiency of the printed holograms is relatively low, with only approximately $2\%$ of the incident optical power distributed between the $-1$ and $+1$ diffraction orders, where the Laguerre–Gaussian beams carrying topological charges $\ell$ and $-\ell$, respectively, are generated. The remaining optical power is distributed among the zeroth and higher diffraction orders and is therefore unavailable for OAM transfer and optical trapping. Nevertheless, the power contained in the first diffraction orders is sufficient to induce stable rotational motion in trapped microparticles. Consequently, and as confirmed experimentally, microparticles trapped in the $+1$ diffraction order exhibit clockwise rotation for positive topological charges $+\ell$ and counterclockwise rotation for negative $\ell$. In contrast, particles trapped in the $-1$ diffraction order rotate in the opposite sense.

%\section{Results and discussion}

For all the experiments only the $+1$ diffraction order was used. The  \textbf{Section V of the Supplementary Material} includes representative videos of approximately 5000 frames for each particle size and topological charge, identified by their respective labels.

\begin{figure}
    \centering
    \includegraphics[width=0.9\linewidth]{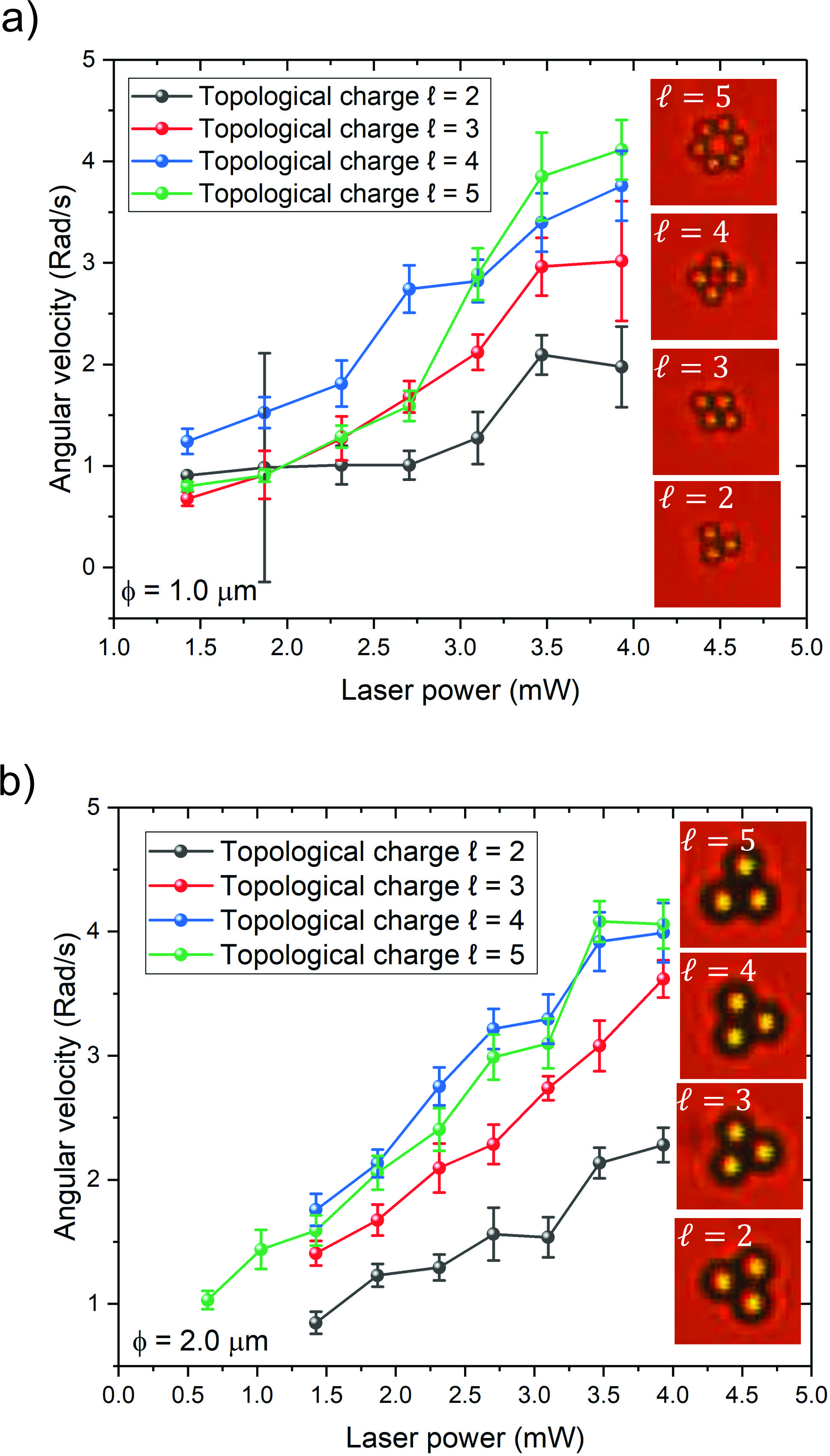}
       \caption{Angular velocity of trapped microparticles, which orbit the bright region of the optical vortex in the trapping plane of the optical tweezers system. a) For microparticles of $1\ \mu\mathrm{m}$ in diameter, and b) for microparticles of $2\ \mu\mathrm{m}$ in diameter. The right vertical insets correspond to a frame of the videos included for each particle size and beam's topological charge.}
    \label{fig:results1}
\end{figure}

The results for the $1\ \mu\mathrm{m}$ particles are summarised in Fig. \ref{fig:results1}a), which shows the angular velocity of the microparticles as a function of laser power, for the various topological charges. As expected, the angular velocity of the microparticles increases with an increase in the laser power. In a similar way, it increases with the absolute value of the topological charge, except for $\ell=5$, where some measured angular velocity values do not follow this trend, which  is attributed to possible collective dynamics.  Here, the error bars were computed as the standard deviation of the average angular velocity, calculated from measurements per complete revolution. For each topological charge and power, three videos of 5000 frames approximately were recorded, and since each particle completed a different number of revolutions, the number of data points used varied in each case. The right vertical insets correspond to a frame of the recorded videos for this particle size and topological charges, which are in the \textbf{section V of the Supplementary Material} and included as video 2 to video 5.

Similar results were obtained for the $2\ \mu\mathrm{m}$ particles, which are shown in Fig. \ref{fig:results1}b), nonetheless some relevant differences were observed. First, the number of trapped particles differs between the two cases, which is related to the ratio between the size of the light distribution (beam) and the size of the particles. In the case of smaller particles, the number of trapped particles increases as the beam size increases, since there is more space available for them to be accommodated. In contrast, for larger particles, this number remains constant because the increase in beam size is not sufficient to accommodate additional particles. In a similar way to the previous case, the right vertical insets correspond to a frame of the recorded videos for this particle size and topological charges, which are in \textbf{Section V of the Supplementary Material} and included as video 6 to video 9.

It is worth emphasising that an accurate computation of the angular velocity requires a full analysis of additional effects, such as fluid viscosity. This is because, in the optical tweezers regime, the particle motion is overdamped, that is, the Reynolds number is much smaller than unity ($Re \ll 1$), and viscous forces dominate over inertial contributions. The angular velocity is therefore set by the balance between the optical torque exerted by the doughnut-shaped beam and the viscous drag torque imposed by the surrounding liquid and any nearby particles. In the case of multiple rotating particles, several physical mechanisms compete and become coupled. For instance, each particle induces a local fluid flow that increases the effective drag experienced by the others, thereby reducing their rotation rates. In addition, collective hydrodynamic interactions can give rise to instabilities, where small differences in angular position are amplified over time, occasionally leading to the formation of transient ``particle chain''. A full quantitative description of these many-body effects would require dedicated simulations based on Brownian dynamics, which lies beyond the scope of the present work.

It was noted that trapping occurs in the $+1$ diffraction order and that only $1\%$ of the laser power reaches this order. In these experiments, very little power was required to set the microparticles in motion. Thresholds of $1.4\ \mathrm{mW}$ were observed for microparticles of $1\ \mu\mathrm{m}$ in diameter and $0.6\ \mathrm{mW}$ for microparticles of $2\ \mu\mathrm{m}$ in diameter. Furthermore, it was observed that larger microparticles are more easily set into motion, since, under the experimental conditions, the trap stiffness of the optical tweezers system decreases with increasing particle size. These results indicate that structuring the beam enhances the effectiveness of optical trapping, as multiple microparticles can be set into motion using lower powers than those typically required for a single optical tweezer.

%\section{Conclusion}
In this work, a low-cost setup of holographic optical tweezers was demonstrated to trap and manipulate microparticles. By harnessing the OAM of the structured light fields, rotational motion was imparted to the trapped microparticles. This result confirms the transfer of the light's angular momentum to the mechanical angular momentum of microparticles.

The results demonstrate an economically viable method for generating optical vortices and implementing holographic optical tweezers. The ability to utilise readily available and inexpensive materials such as printed acetate sheets for OAM-based micromanipulation opens new possibilities for research and applications in fields such as microfluidics, biophysics, and micro-robotics, where controlled rotation of microscopic objects is essential. In addition, the approach naturally enables the generation of pairs of optical vortices with opposite topological charge.

Although binary holograms printed on acetate sheets are not the only low-cost holograms, it was found that they offer several advantages over phase holograms fabricated using other methods, such as those developed on photographic film. First, they provide improved reproducibility, since they do not require precise exposure times to different chemicals, which could lead to variations in the final hologram performance. Another advantage is their durability, as they are not affected by changes in temperature or humidity, as in the case of gelatin in photographic film, which can swell and alter the microscopic pattern.

\section*{Acknowledgement}

The author thanks E. Sarmiento-Gómez for providing essential optical components for the experimental setup and SECIHTI for its support through the project CBF-2025-I-1804.

\bibliography{references.bib}
\end{document}

\title{Supplementary Material: Robust Orbital Angular Momentum Transfer Using Low-Cost Diffractive Optics}

\maketitle

\section{Orbital angular momentum in optical tweezers}

Angular momentum is one of the fundamental conserved quantities in physics, along with energy and linear momentum. A rigid body can exhibit two kinds of rotational motion: it can spin around its axis, and it can rotate around a distant point. For example, the daily rotation of the Earth gives rise to spin angular momentum, while its motion around the Sun generates orbital angular momentum. In optics, it is well known that when an atom absorbs a circularly polarized photon, it acquires angular momentum of $\pm \hbar$ in the propagation direction. More generally, light carries two distinct forms of angular momentum: Spin Angular Momentum (SAM) and Orbital Angular Momentum (OAM). The SAM, linked to circular polarization, was predicted theoretically by Poynting in 1909 \cite{Poynting1909}; however, its experimental verification remained elusive until 1936, when Beth measured the corresponding angular momentum transfer using a torsion balance and a birefringent medium \cite{1936_Beth}. By contrast, OAM is associated with the spatial structure of the optical wavefront (specifically beams exhibiting azimuthally varying phase profiles) and was formally introduced in 1992 by Allen \textit{et al}., who showed that Laguerre-Gaussian (LG) modes carry a well-defined and quantised amount of OAM \cite{1992_Allen}.

In the context of optical tweezers, both forms of angular momentum can be transferred to trapped microparticles, giving rise not only to linear trapping forces but also to optical torques. While SAM induces particle rotation through polarisation-dependent spin transfer, OAM enables controlled mechanical motion driven by the helical phase structure of the beam. Consequently, a circularly polarised Gaussian beam can transfer SAM to birefringent \cite{1998_Friese} or absorbing microparticles \cite{1996_Friese,1998_Friese_torque}. Conversely, OAM manifests as a local transverse momentum component that can drive orbital motion around the beam axis. This distinction gives rise to the concepts of intrinsic OAM, associated with the optical vortex itself, while extrinsic OAM depends on the position of the particle relative to the beam axis \cite{2002_Oneil_intrinsic}. For transparent spherical particles, extrinsic OAM induces orbital rotation, whereas intrinsic OAM can drive axial rotation in asymmetric objects. In this context, in 2001 Paterson \textit{et al.} demonstrated axial rotation of asymmetric transparent particles, which included glass rods and chromosomes, induced by OAM transfer \cite{2001_Paterson}. Crucially, the simultaneous transfer of SAM and OAM can be achieved using circularly polarised LG beams \cite{1995_He_optical,1995_He_direct}. Collectively, these developments have expanded optical trapping beyond static confinement, enabling optically driven micromotors and controlled rotational dynamics at the microscale \cite{galajda2002rotors,padgett2011tweezers}.

As discussed previously, Laguerre–Gaussian (LG) beams constitute one of the most important classes of structured light fields carrying OAM. They form an infinite, orthogonal set of solutions to the paraxial wave equation expressed in cylindrical coordinates $(\rho,\varphi, z)$. Mathematically, LG beams are described by \cite{Siegman},
\begin{eqnarray}
\label{eq:LG_complexfield}
       U(\rho,\varphi,z)&=&\frac{w_0}{w(z)}\sqrt{\frac{2p!}{\pi(|\ell|+p)!}} \left( \frac{\sqrt{2}\rho}{w(z)}\right)^{|\ell|} L_p^{|\ell|} \left[2\left(\frac{\rho}{w(z)}\right)^2\right]\\ \nonumber
  && \exp{[-i(2p+|\ell|+1)\zeta(z)]}\exp\left[-\left({\frac{\rho}{w(z)}}\right)^2\right]\\ \nonumber
  &&\exp{\left[-\frac{ik\rho^2}{2R(z)}\right]}\exp(i\ell\varphi),
\end{eqnarray}

\noindent here, $\ell\in\mathbb{Z}$, known as the topological charge, specifies the number of $2\pi$ phase windings around the optical axis, whereas $p \in \mathbb{N}$ is the radial index. In addition, $w_0$ denotes the beam waist at the focus, $w(z)=w_0\sqrt{1+\left(\frac{z}{z_R}\right)^2}$ the beam radius upon propagation, $z_R = \pi w_0^2/\lambda$ the Rayleigh range. Furthermore, $L_p^{|\ell|}[\cdot]$ are the associated Laguerre polynomials of order $(|\ell|,p)$, $R(z)=z\sqrt{1+\left(\frac{z}{z_R}\right)^2}$ represents the radius of curvature of the wavefront, $\zeta(z)=\arctan{\left(\frac{z}{z_R}\right)}$ is the Gouy phase, $k=2\pi/\lambda$ and $\lambda$ is the wavelength. Importantly, the helical phase term $\exp(i\ell \varphi)$ endows the beam with an orbital angular momentum of $\ell\hbar$ per photon. Of particular interest is the subset of LG modes with $p=0$, often referred to as optical vortices. These modes exhibit an on-axis phase singularity that results in the characteristic doughnut-shaped intensity profile. This null intensity at the origin ($\rho=0$) arises from the factor $\left(\frac{\sqrt{2}\rho}{w(z)}\right)^{|\ell|}$, which enforces the central zero and increases the ring radius with $|\ell|$. 

Although SAM and OAM arise from different physical origins, for a light-absorbing particle they are mechanically equivalent, as both contribute to an optical torque capable of inducing rotation. Simpson \textit{et al.} \cite{1997_Simpson} experimentally demonstrated that, from a mechanical standpoint, SAM (associated with polarisation) and OAM (associated with the beam’s wavefront) contribute additively to the total angular momentum transferred to a particle. Accordingly, the total angular momentum carried by a tightly focused LG beam is 
\begin{equation}
   J= \left[\ell+\sigma_z+\sigma_z\left(\frac{2kz_R}{2p+\ell+1}+1\right)^{-1}\right]\hbar,
\end{equation}
where $\sigma_z=\pm1$ denotes the spin angular momentum corresponding to right- or left-handed circular polarisation. The last term accounts for spin-orbit coupling effects arising from tight focusing. In the paraxial limit $kz_R \gg 1$, this expression reduces to $J=(\ell+\sigma_z)\hbar$ per photon.

For linearly polarised light ($\sigma_z=0$), even under strong focusing conditions such as those in optical trapping, the total angular momentum transferred per second is given by \cite{1995_He_direct} 
\begin{equation}
   \tau = \frac{P}{\omega}\ell,
\end{equation}
where $P$ is the laser power and $\omega$ is the angular frequency of the light beam.

\section{Binary hologram encoding}
In the past, several cost-effective beam-shaping strategies have been explored, including photographic reduction \cite{1966_Brown,1995_He_optical} photographic films, \cite{2007_Mariscal} transparency sheets, \cite{2013_Kumar} and even repurposed LCD panels from commercial projectors \cite{2012_Huang}, their practical implementation is often hampered by fabrication complexity, limited reproducibility, and susceptibility to thermal or radiation-induced degradation. In contrast, designing and printing binary holograms on acetate sheets provides a simple, durable, and mass-reproducible platform for the high-fidelity generation of optical beams carrying OAM \cite{2023_Torres}, which, as will be demonstrated later, can reliably transfer OAM from light to microparticles. 

By exploiting spatial variations in transmittance, binary amplitude masks are designed to generate a first diffraction order that faithfully reconstructs the desired intensity and phase distribution of a target optical field \cite{1966_Brown,1979_Lee}. In our implementation, the complex LG field of Eq.~\ref{eq:LG_complexfield} at z=0 is encoded into the binary transmittance function,
\begin{equation}
    T(\rho,\varphi)=\frac{1}{2}+\frac{1}{2}\sgn\{\cos[p(\rho,\varphi)]+\cos[q(\rho,\varphi)]\}, 
\end{equation}
with
\begin{eqnarray}
    p(\rho,\varphi) &=& \arcsin\!\left[\dfrac{|U_{\ell,p}(\rho,\varphi)|}{\max\{ |U_{\ell,p}(\rho,\varphi)|\}}\right],\\
    q(\rho,\varphi) &=& \phi(\rho,\varphi)+2\pi(\nu x+\eta y),
\end{eqnarray}
where $\mathrm{sgn}\{\cdot\}$ denotes the sign function and $\max\{\cdot\}$ the maximum amplitude of the field. The term $\phi(\rho,\varphi)$ is the phase of the LG beam, and $2\pi(\nu x + \eta y)$ corresponds to a linear grating with spatial frequencies $(\nu, \eta)$, that cleanly separate diffraction orders, ensuring that the encoded LG field appears with high fidelity in the first diffracted order. The transmittance function was printed on a transparent acetate sheet using a high-resolution printer.

\section{Experimental methodology}

Optical trapping experiments were conducted using structured light beams, more precisely, Laguerre-Gaussian profile on samples containing spherical polystyrene microparticles immersed in water. The sample consisted of a mixture of water with polystyrene particles of different diameters, whose refractive index is 1.59. Samples were prepared with particles of $1\ \mu\mathrm{m}$ (PS-R-L2910-1, microparticles GmbH) and $2\ \mu\mathrm{m}$ (78452-5ML-F, Sigma Aldrich) in diameter at concentrations of $5.3\times 10^{-4}\ \mathrm{spheres/\mu m^3}$ and $2.2\times 10^{-4}\ \mathrm{spheres/\mu m^3}$. Since these are dielectric particles, they are attracted toward regions of higher intensity due to optical gradient forces.

\begin{figure}
    \centering
    \includegraphics[width=0.7\linewidth]{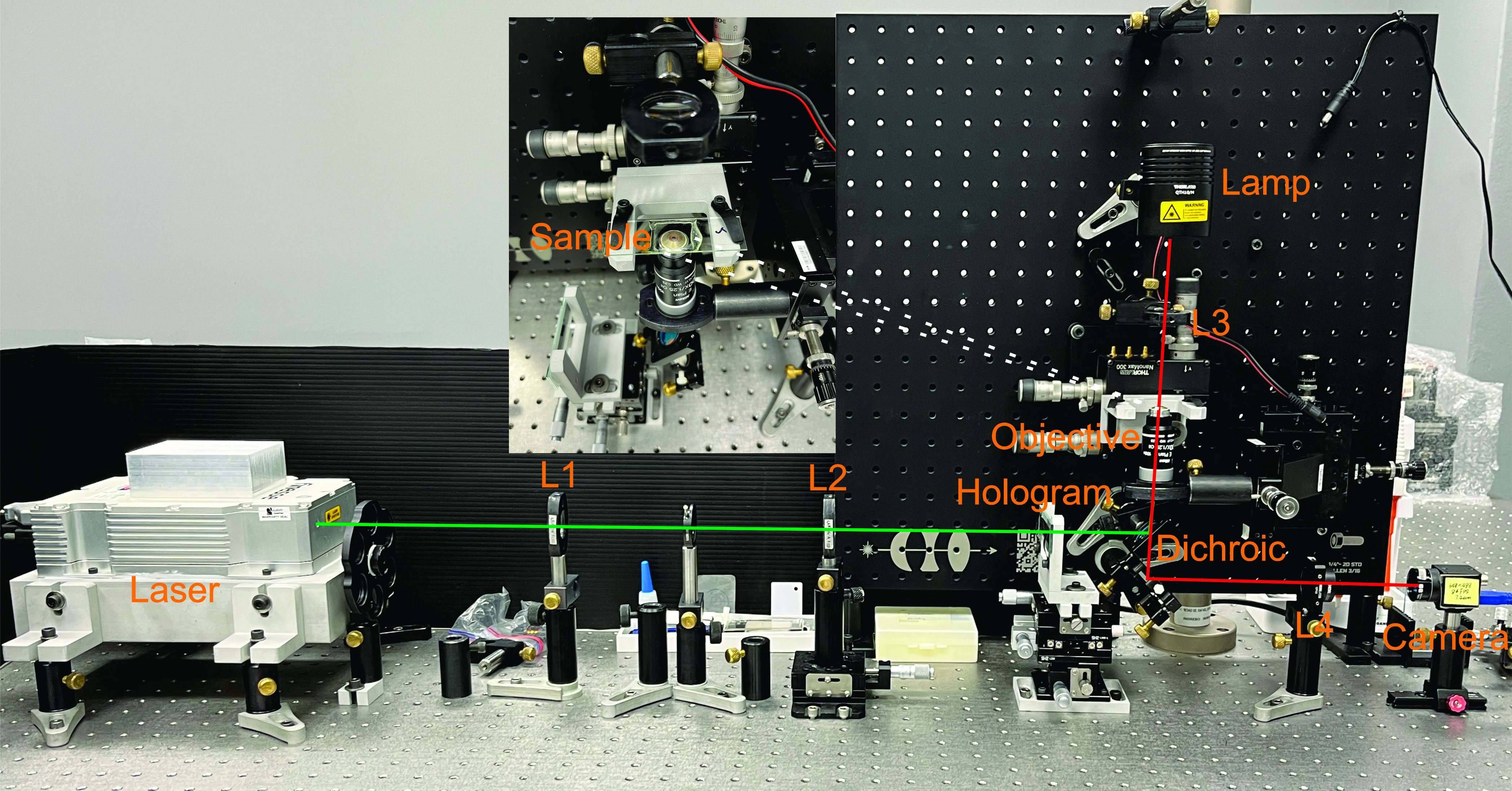}
    \caption{Photograph of the experimental setup, it shows the basic components of an optical tweezers system, as well as the low-cost hologram. The inset in the upper center indicates the sample position.}
    \label{fig:setup}
\end{figure}

A schematic representation of the experimental setup is shown in Fig. \ref{fig:setup}. The system relies on a Gaussian trapping beam generated by a continuous-wave laser operating at 532 nm with a maximum output power of 4 W (finesse, Laser Quantum). The laser power was tuned from 0 to 400 mW using neutral-density filters mounted on a filter-wheel station (FW1A, Thorlabs), together with fine adjustments from the laser controller. The beam was subsequently expanded using a telescope composed of two plano-convex lenses, L1 and L2, with focal lengths of 100 mm and 125 mm and reflected afterwards by a dichroic mirror (DMPL605, Thorlabs) towards the optical tweezers system. Finally, the beam was tightly focused onto the sample plane using a 100x microscope objective (E Plan 100x/1.25, Nikon, NA= 1.25) to achieve stable trapping. To image the trapping region, the sample was illuminated with a halogen lamp (QTH10/M, Thorlabs). A condenser lens L3 (LA1131-A, Thorlabs) formed the corresponding image through the same microscope objective used for trapping. The dichroic mirror enabled imaging of the sample with red light from the halogen lamp. More specifically, the microscope objective, together with an eyepiece lens L4 (AC254-100-A-ML, Thorlabs), relayed the sample plane to a CCD camera (BFLY-U3-03S2C-CS, FLIR), which records videos at a rate of 84 frames per second. To avoid backscattered light towards the camera, a bandpass filter (FELH0600, Thorlabs) was placed in front of the camera.

Laguerre-Gauss vortex beams were generated using acetate-printed holograms with topological charges between $\ell=2$ and $\ell=5$. These beams, with a doughnut-shaped intensity profile, were used to trap and rotate microparticles suspended in water. Experiments were conducted by varying the laser power and the topological charge. The particle dynamics were recorded using a CCD camera, and rotation was analyzed from video tracking. 

\section{Particle tracking and data analysis}

The generated Laguerre-Gauss beams transfer orbital angular momentum to the trapped microparticles, causing them to rotate. Despite the low efficiency of the printed holograms (only about 2\% of the power is concentrated in the useful diffraction orders), the beams produce a stable rotation whose angular velocity increases linearly with the topological charge. Furthermore, the direction of particle rotation depends on the sign of the topological charge and the diffraction order used. 

\begin{figure*}
    \centering
    \includegraphics[width=0.7\linewidth]{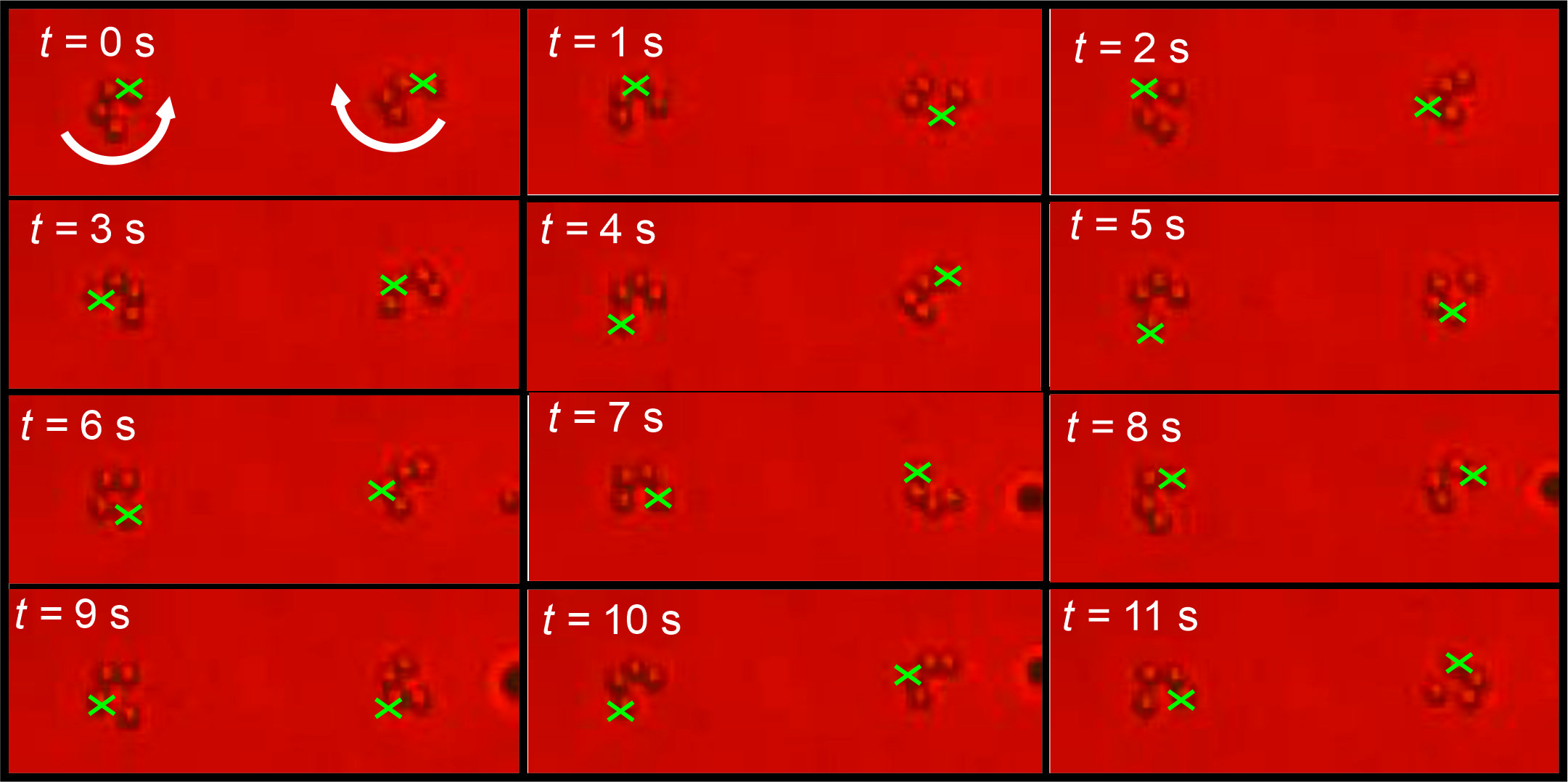}
    \caption{Sequence of frames extracted from video 1, showing the movement of spherical particles trapped in the $+1$ (right) and $-1$ (left) orders of the doughnut-shaped beam with topological charge $\ell= 4$. Each tracked particle was identified with an ``x''.} 
    \label{fig:frames2}
\end{figure*}

This behaviour is illustrated in Fig. \ref{fig:frames2}, which shows a sequence of frames extracted from a 11-second video section (the complete video is provided in the {\bf Supplementary Material} as video 1), where it is observed that the microparticles trapped in the diffraction order $+1$ rotate clockwise for positive topological charges and rotate counterclockwise for negative topological charges, while the microparticles trapped in the diffraction order $-1$ rotate in the opposite direction. Four spherical microparticles with a diameter of $1\ \mu\mathrm{m}$ are trapped in the $-1$ (left) diffraction order and another four in $+1$ (right) diffraction order of the doughnut-shaped beam. The two diffraction orders effectively behave as optical vortices of equal magnitude but opposite topological charge and therefore, each cluster rotate in opposite directions, anticlockwise and clockwise, respectively, as indicated by the circular arrows. Importantly, the beams generated in this way are suitable for realising optically driven microfluidic pumps, where the separation of the microrotors can be controlled by adjusting the grating period \cite{2006_Leach}. In addition, this configuration naturally enables the generation of a vector beam through a nonseparable superposition of the two vortices carrying orthogonal polarisation states. Such beams are of particular interest in optical tweezers, as they can enhance both transverse and longitudinal trapping forces \cite{Bhebhe2018vector,2018_Rosales}. 

\begin{figure}
    \centering
    \includegraphics[width=0.6\linewidth]{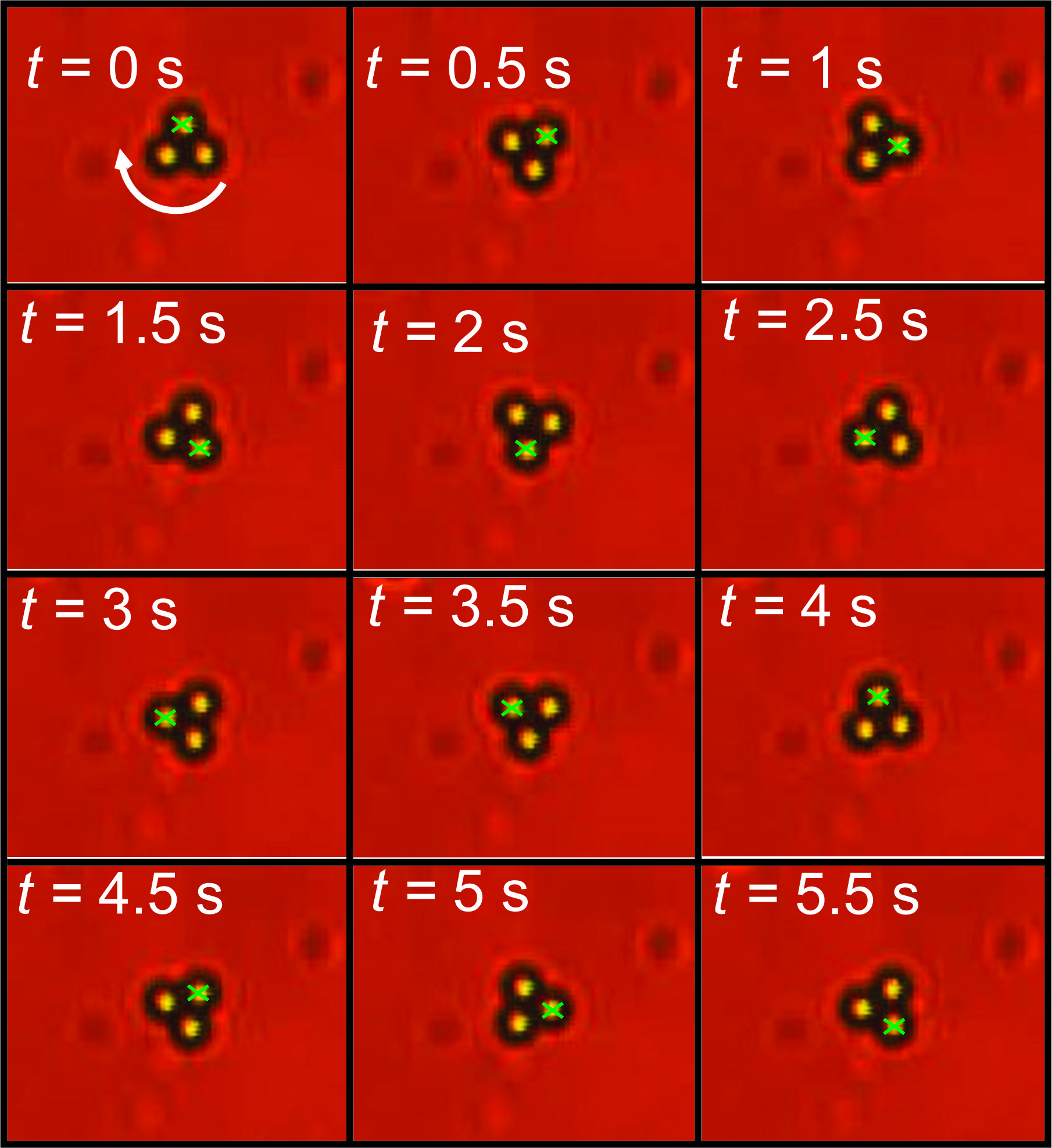}
    \caption{Sequence of frames, extracted from video 6, showing the uniform circular motion of particles trapped in the doughnut-shaped structured optical tweezers system with topological charge $\ell=5$.} % and 142 mW of optical power, measured at the laser output
    \label{fig:frames}
\end{figure}

Additional experiments were performed for different power levels and topological charges ranging from $\ell=2$ to $\ell=5$, with the resulting particle dynamics recorded on video for subsequent analysis. The data processing consisted of extracting the video information frame by frame, identifying the centroid of a selected microparticle, and tracking its trajectory throughout the recording. This procedure allows to determine the time required for the particle to complete one full revolution and, consequently, to calculate its angular velocity. By way of example, Fig. \ref{fig:frames} shows a sequence of frames extracted from one of the videos (included in the {\bf Supplementary Material} as video 6), illustrating the tracking of a particle, marked with an ``x" for reference, that demonstrates the uniform circular motion of particles trapped in a LG beam of topological charge $\ell=5$. 

The analysis also included the development of a specific algorithm to calculate the radius of the circular path described by the particles rotating around the optical vortex. The algorithm requires an image obtained by averaging all video frames in the region corresponding to one of the diffraction orders. The algorithm then quantifies the intensity of each pixel in the radial direction, swept over an angle of $2\pi$ to cover the entire image, and averages the values. The resulting data are fitted to a Gaussian distribution to identify the maximum intensity, which coincides with the radius of the trajectory.

\section{Videos}
The \href{https://drive.google.com/drive/folders/1t0OZDCXzkjHaYn8iJyOv74liU5lsFsM3?usp=sharing}{link} contains a video labeled as ``video 1'', which shows the trapping plane of the optical tweezers system and demonstrates the orbital motion of $1\ \mu\mathrm{m}$ microparticles trapped in diffraction orders $+1$ (right) and $-1$ (left). Videos 2-5 show diffraction order $+1$ with $1\ \mu\mathrm{m}$ microparticles for topological charges from 2 to 5 and constant laser power. Videos 6-9 show diffraction order $+1$ with $2\ \mu\mathrm{m}$ microparticles for topological charges from 2 to 5 and constant laser power. Additionally, in video 10 it can be seen the orbital motion of microparticles confined by higher diffraction orders.

\bibliography{references.bib}